\documentclass[a4paper]{jpconf}

\usepackage{graphicx}

\relax 
\bibstyle{elsarticle-num}
\bibdata{mybibfile}
\begin{document}

\sloppy 

\title{Study of external electron beam injection into proton driven plasma wakefields for AWAKE Run2}



\author{L. Verra$^{1,2,3}$, E. Gschwendtner$^{1}$ and P. Muggli$^{2}$}
\address{$^1$ CERN, Geneva, Switzerland}
\address{$^2$ Max-Planck Institute for Physics, Munich, Germany}
\address{$^3$ Technical University Munich, Munich, Germany}

\ead{livio.verra@cern.ch}

\begin{abstract}

We describe an external electron injection scheme for the AWAKE experiment. We use scattering in two foils, that are necessary as vacuum window and laser beam dump, to decrease the betatron function of the incoming electron beam for injection and matching into plasma wakefields driven by a self-modulated proton bunch. We show that, for a total aluminum foil thickness of $\sim 280\, \mu\textnormal{m}$, multiple Coulomb scattering increases the beam emittance by a factor of $\sim 10$ and decreases the betatron function by a factor of $\sim 3$. The plasma in the accelerator is created by a ionizing laser pulse, counter-propagating with respect to the electron beam. This allows for the electron bunch to enter the plasma through an "infinitely" sharp vapor-plasma boundary, away from the foils. 

\end{abstract}

\section{Introduction}

\begin{figure}[!h]
\centering
\includegraphics[scale=0.21]{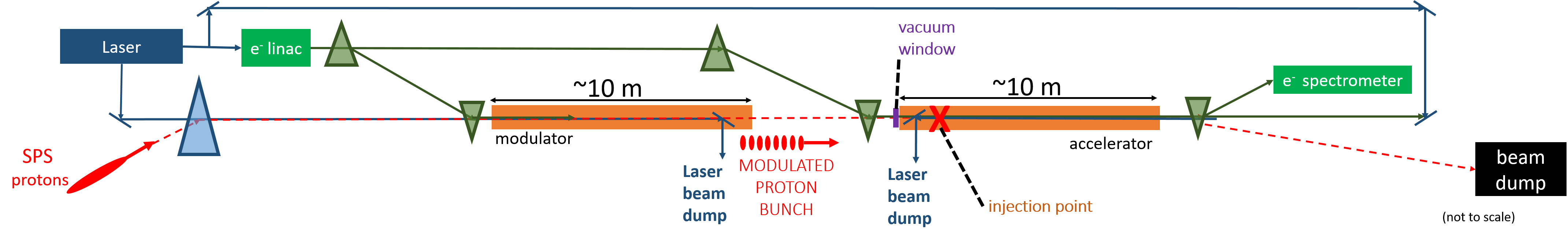}
\caption{Schematic drawing of the AWAKE Run2 setup.}
\label{fig:setup}
\end{figure}

During its first experimental run (2016-2018), AWAKE (the Advanced WAKEfield experiment) $\cite{PATRIC:READINESS}$ reached two important milestones: the demonstration of the seeded self-modulation of the $400 \,\textnormal{GeV/c}$ proton bunch delivered by the CERN Super Proton Synchrotron $\cite{MARLENE:SSM}\cite{KARL:SSM}$, and the acceleration of externally injected electrons from $19 \,\textnormal{MeV}$ up to $2\,\textnormal{GeV}$ $\cite{NATURE}$. The goal of the second run is to accelerate a $165 \,\textnormal{MeV}$ electron bunch while preserving its quality. For AWAKE Run2 $\cite{PATRIC:RUN2}$ we plan to use two separated plasma sources: one dedicated to the self-modulation of the proton bunch (seeded by an electron bunch) and one to the electron acceleration (see {Figure $\ref{fig:setup}$}). 
\par A short vacuum gap (with length of $\sim 30 \,\textnormal{cm}$) separates the two sources, and aluminum windows confine the rubidium vapor to the sources. In the gap region, the electron beam trajectory merges with the proton beam one, so as to inject the witness bunch on axis into the wakefields. A laser pulse ($\sigma_{\tau} = 120 \,\textnormal{fs}$, $E <450 \textnormal{mJ}$, $\lambda= 780 \,\textnormal{nm}$) is split to ionize separately the rubidium vapor in the two sources. In particular, the pulse ionizing the accelerating section counter-propagates with respect to the beams. Thus, a laser beam dump protects the vacuum window in each source. Hence, the proton and electron beams cross these aluminum foils upstream the injection point. While the effect on the proton beam optical properties is negligible, the electron beam is strongly affected by scattering in the material. In this paper we study the incoming electron beam parameters to achieve matching to the plasma focusing force.

\section{Electron beam injection}
\subsection{Blowout, beam loading, beam matching}
\par The final goal of the AWAKE experiment is to provide an electron beam suitable for applications to high-energy physics (fixed target or electron-proton collision experiments). To do this, it is necessary that the electron bunch carries a high charge ($>100\,\textnormal{pC}$), that the normalized emittance is sufficiently low ($10 - 20 \,\textnormal{mm}\cdot\textnormal{mrad}$), and that the final energy spread is kept at, or below, the $\%-$level.
\par The bunch must be injected in the accelerating and focusing phase of the wakefields. Therefore, the electron bunch length must be much shorter than a quarter of the plasma electron wavelength $\lambda_{pe} = 2\pi c/\omega_{pe}$, where $c$ is the speed of light and ${\omega_{pe}=\sqrt{n_{pe} e^2/\epsilon_0 m_e}}$ is the angular plasma electron frequency ($n_{pe}$ is the plasma electron density, $e$ is the elementary charge, $\epsilon_0$ is the vacuum permittivity, $m_e$ is the electron mass). This would insure high efficiency of the capture process. At the baseline plasma electron density ($n_{pe} = 7\cdot 10^{14}\,\textnormal{cm\textsuperscript{-3}}$), $\lambda_{pe}\sim 1.2 \,\textnormal{mm}$.
\par According to Liouville's theorem, the incoming emittance is preserved if the transverse focusing force acting on the witness beam increases linearly with the distance from the axis. This is achieved by fully blowing out the plasma electrons from the plasma cavity $\cite{ROSENWEIG:BLOWOUT}$: the system enters in the so-called blowout, non-linear regime. In this scenario, the focusing force generated by the pure, uniform density ion column is radially linear, therefore the electron bunch slice emittance can be conserved. Simulations $\cite{VERONICA:BLOWOUT}$ show that, even though the AWAKE proton microbunches generate plasma wakefields only in the quasilinear regime ($\delta n_{pe} \le n_{pe}$), an intense enough electron bunch ($n_{eb} >35 n_{pe}$, with $n_{eb}$ the charge density of the electron bunch, for a bunch length of $60\, \mu\textnormal{m}$) can expel all the residual plasma electrons from the propagation axis, leaving an ion column behind. 
\par In order to accelerate this bunch with a low energy spread, it is necessary to flatten the longitudinal wakefield amplitude along the bunch, so that most of the witness bunch particles experience the same accelerating gradient. This is possible with beam loading $\cite{KATSOULEAS:LOADING}$: the witness bunch is positioned such that its own wakefields, superimposed to the wakefields driven by the proton bunch train, make the accelerating field approximately constant along the witness bunch.
%
%

\par To maintain blowout and beam loading, the electron bunch charge density may not oscillate while propagating along the plasma. This is satisfied by matching the electron beam to the plasma ion column focusing force. When the beam is injected into the wakefields, its transverse size $\sigma$ follows the envelope equation:
\begin{equation}
    \sigma''(z) + (K^2_{\beta} -\frac{\epsilon^2_g}{\sigma^4(z)})\sigma(z)=0,
\label{eq:envelope}
\end{equation}
where $K_\beta=\frac{\omega_{pe}}{c\sqrt{2\gamma}}$ is the focusing term of the ion column ($\gamma$ is the Lorentz factor), and the $\frac{\epsilon^2_g}{\sigma^4(z)}$ term describes the divergence of the beam due to its geometric emittance $\epsilon_g$. The beam is matched to the plasma (and therefore its envelope does not oscillate along the plasma) when it is injected at the waist ($\sigma'(z_{inj}) = 0$, where $z_{inj}$ is the longitudinal position of the injection point, i.e. the plasma entrance) and the term in parenthesis in Equation $\ref{eq:envelope}$ vanishes, i.e. the focusing force exactly balances the divergence of the beam. Satisfying these conditions yields:
\begin{equation}
    \beta^{*} = \sqrt{\frac{2 \epsilon_0 m_e c^2 \gamma}{n_{pe} e^2}},
\label{eq:matching_beta}
\end{equation}
where $\beta^{*} = \sigma^{*} / \epsilon_g$ is the betatron function of the electron beam at the injection point (therefore, at the beam waist). Thus, Equation $\ref{eq:matching_beta}$ shows that the matching does not depend directly on the emittance or the size of the electron beam, but only on the value of $\beta^{*}$. 
\subsection{Electron beam injection and matching in the AWAKE experiment }
\par
\begin{figure}[h!]
\includegraphics[width=19pc]{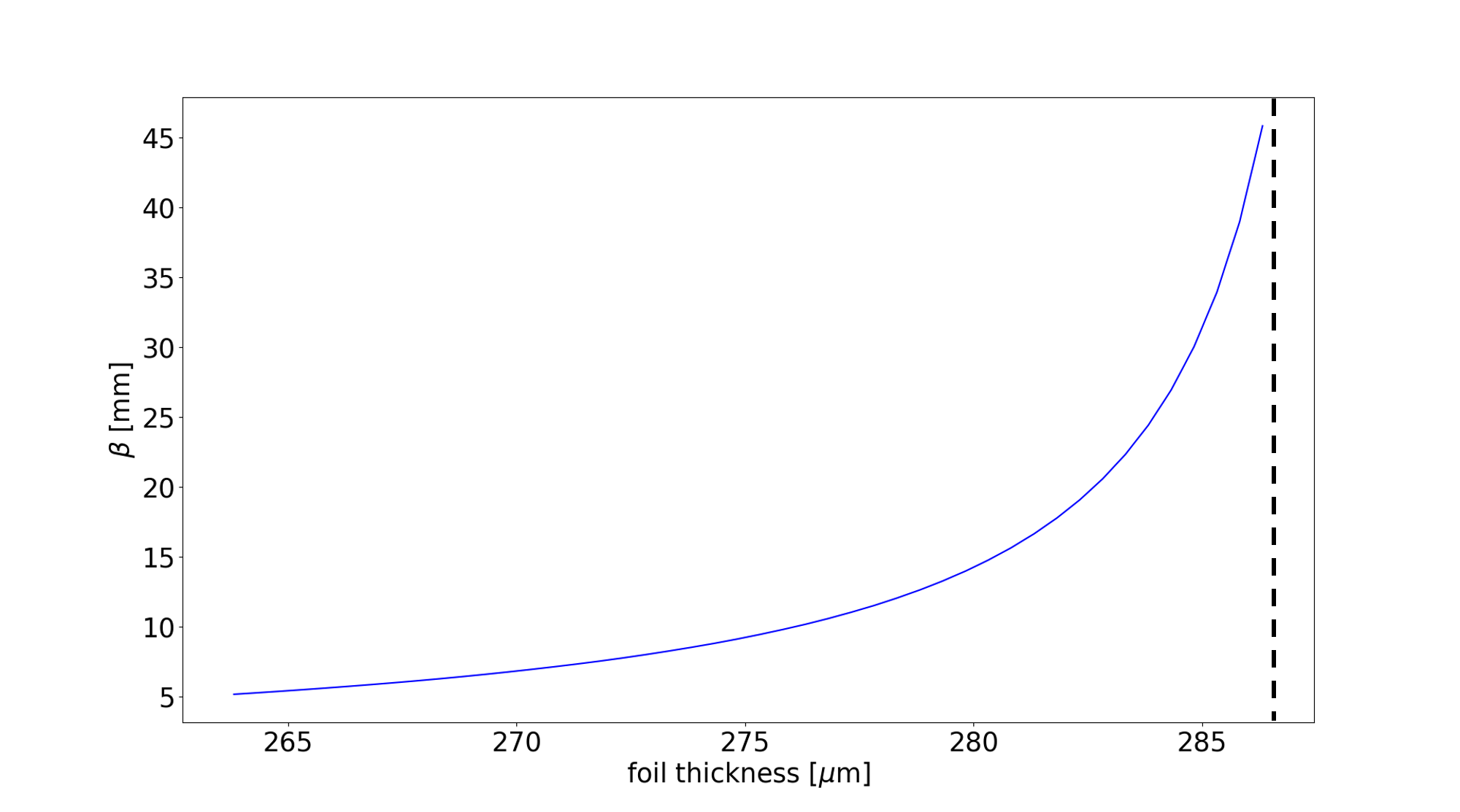}\hspace{2pc}%
\begin{minipage}[b]{16pc}\caption{\label{fig:parameters}Incoming betatron function $\beta^{*}_{in}$ as a function of the foil thickness required to reach the goal parameters at the injection point.}
\end{minipage}
\end{figure}

 For the AWAKE Run2 baseline parameters, the electron beam energy is 165 MeV and ${n_{pe}=7\cdot 10^{14} \,\textnormal{cm\textsuperscript{-3}}}$, hence ${\beta^{*} = 5.1 \,\textnormal{mm}}$. This is a rather short value that is challenging to produce in the AWAKE geometry, since it requires strong focusing close to the waist location. Increasing this value for the incoming beam is therefore desirable. As mentioned above, the electron beam has to cross two aluminum foils before the injection (a vacuum window and a laser beam dump). We choose aluminum because of the good trade-off between its radiation length ($X_0 \sim 9 \,\textnormal{cm}$) and its mechanical properties. The incoming beam parameters are spoiled because of multiple scattering inside the material $\cite{HIGHLAND:SCATTERING}$: the emittance increases, the betatron function decreases, the position of the waist moves upstream $\cite{REID:SCATTERING}$. Therefore, since the plasma parameters determine $\beta^{*}$ after the foils, we calculate backwards the necessary incoming beam parameters (incoming betatron function $\beta^{*}_{in}$ and position of the waist respect to the laser beam dump position) and the maximum possible foil thickness in order to match the beam with the plasma at the injection point, according to:
\begin{equation}
    \epsilon_{in}^2 - \epsilon_g^2 = \sigma_{f2}^2 \theta_{f2}^2 + \sigma_{f1}^2 \theta_{f1}^2,
\end{equation}

where $\epsilon_{in}, \sigma_{f2}, \theta_{f2}$ and $\epsilon_g, \sigma_{f1}, \theta_{f1}$ are the geometric emittance, transverse beam size, scattering angle at the vacuum window and at the laser beam dump, respectively, and

\begin{equation}
    \beta^{*}_{in} = \frac{\epsilon_{in}}{\epsilon_g - \beta^{*} (\theta_{f1}^2 + \theta_{f2}^2)} \cdot \beta^{*},
\label{eq:beta}
\end{equation}
Figure $\ref{fig:parameters}$ shows the required betatron function to achieve a normalized emittance ${\epsilon_N=\beta\gamma\epsilon_g=20\,\textnormal{mm$\cdot$mrad}}$ ($\beta$ is the ratio of the beam velocity to $c$) and $\beta^{*} = 5.1 \,\textnormal{mm}$ at the injection, assuming an initial normalized emittance of $2 \,\textnormal{mm$\cdot$mrad}$ (nominal value provided by the electron beamline design), as a function of the total amount of material in the beam path. 
\par Note that, as the foils thickness increases, $\beta^{*}_{in}$ has to increase, but it is independent of the distance between the two foils (Equation $\ref{eq:beta}$). The black dashed line indicates the maximum amount of material that can be positioned in the beam path. For a foil thicker than this value, the divergence contribution of the multiple scattering $<\theta^2>$ becomes too large to be compensated by any convergence angle. Therefore, the beam defocuses at the foil exit and cannot be matched to the plasma wakefields. The decrease of the betatron function, as the beam crosses the material, means that the beam waist moves closer to the foil.
\par Hence, the beamline does not have to provide directly $\beta^{*}_{in} = \beta^{*}$, that is very challenging to produce and require to position the last focusing element very close to the injection point. Instead, we exploit the two foils, needed as vacuum window and beam dump, to relax the request on the betatron function.
\par
\begin{figure}[h!]
\includegraphics[width=17pc]{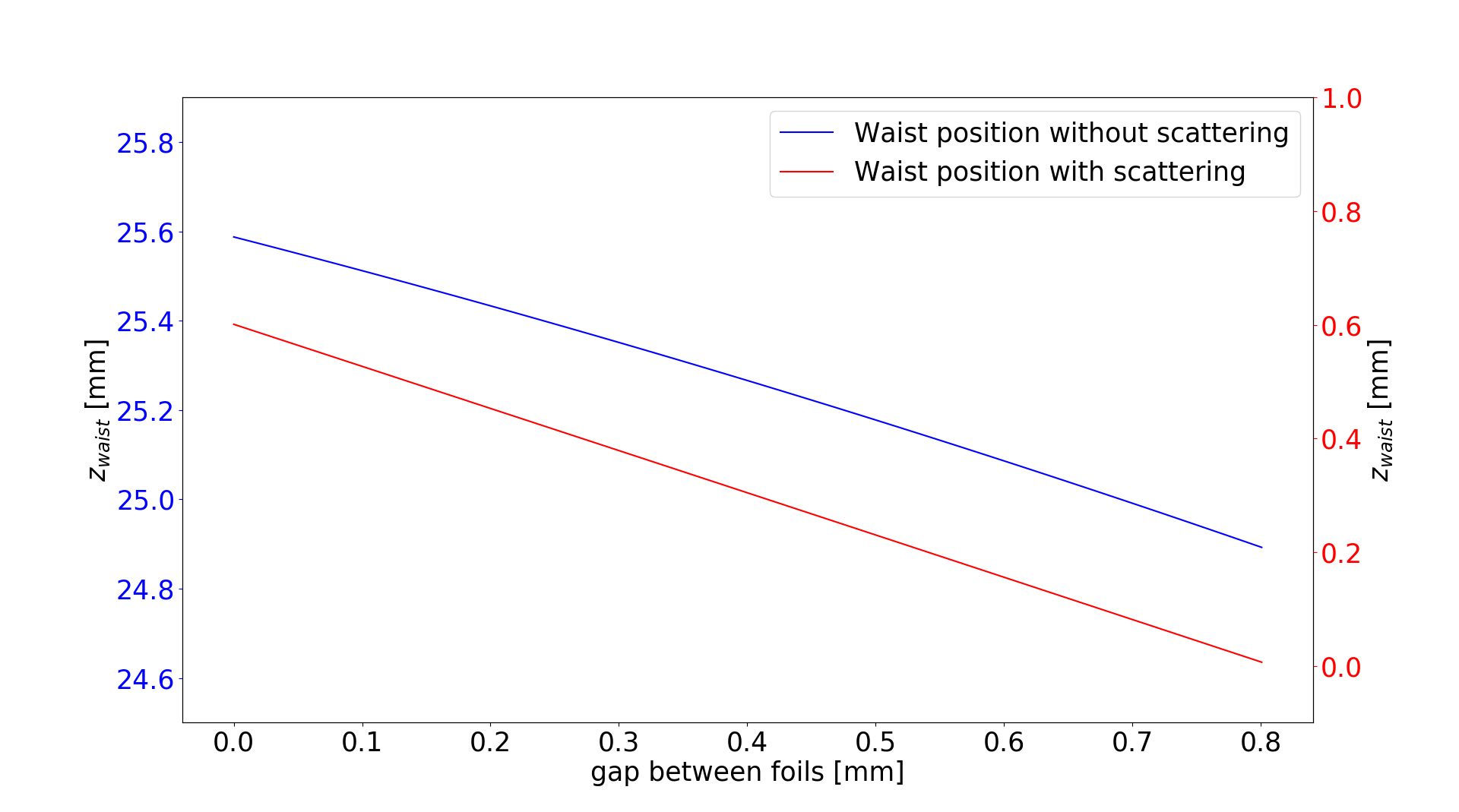}\hspace{2pc}%
\begin{minipage}[b]{17pc}\caption{\label{fig:waist}Position of the waist with (orange line) and without (blue line) scattering, respect to the laser beam dump, as a function of the gap between the two foils.}
\end{minipage}
\end{figure}

 We also calculate the position respect to the laser beam dump where the incoming beam waist needs to be set (without foils and scattering), as a function of the distance between the two foils (Figure $\ref{fig:waist}$, blue line). The upper limit is given by the requirement that the beam does not diverge upstream the laser beam dump. Figure $\ref{fig:waist}$ also shows the final position of the waist with foils and scattering (orange line): this is closer to the foils (i.e. upstream) than without scattering, as expected. Using the calculated betatron function and waist position, we can estimate the maximum distance from the plasma entrance where the last focusing element can be positioned. Considering a quadrupole magnet good field region radius $r\sim 20\,\textnormal{mm}$, the upper limit of the distance for the magnet to accept the whole beam (i.e. $3\sigma < r$) is $\sim 10 \,\textnormal{m}$. Still, in order to keep $\sigma < 10 \,\textnormal{mm}$ inside the magnets, the last focusing element is positioned as closed as possible to the gap region, depending on the beamline bending angle.
%

\par
\begin{figure}[h!]
\centering
\begin{minipage}{17pc}
\includegraphics[width=17pc]{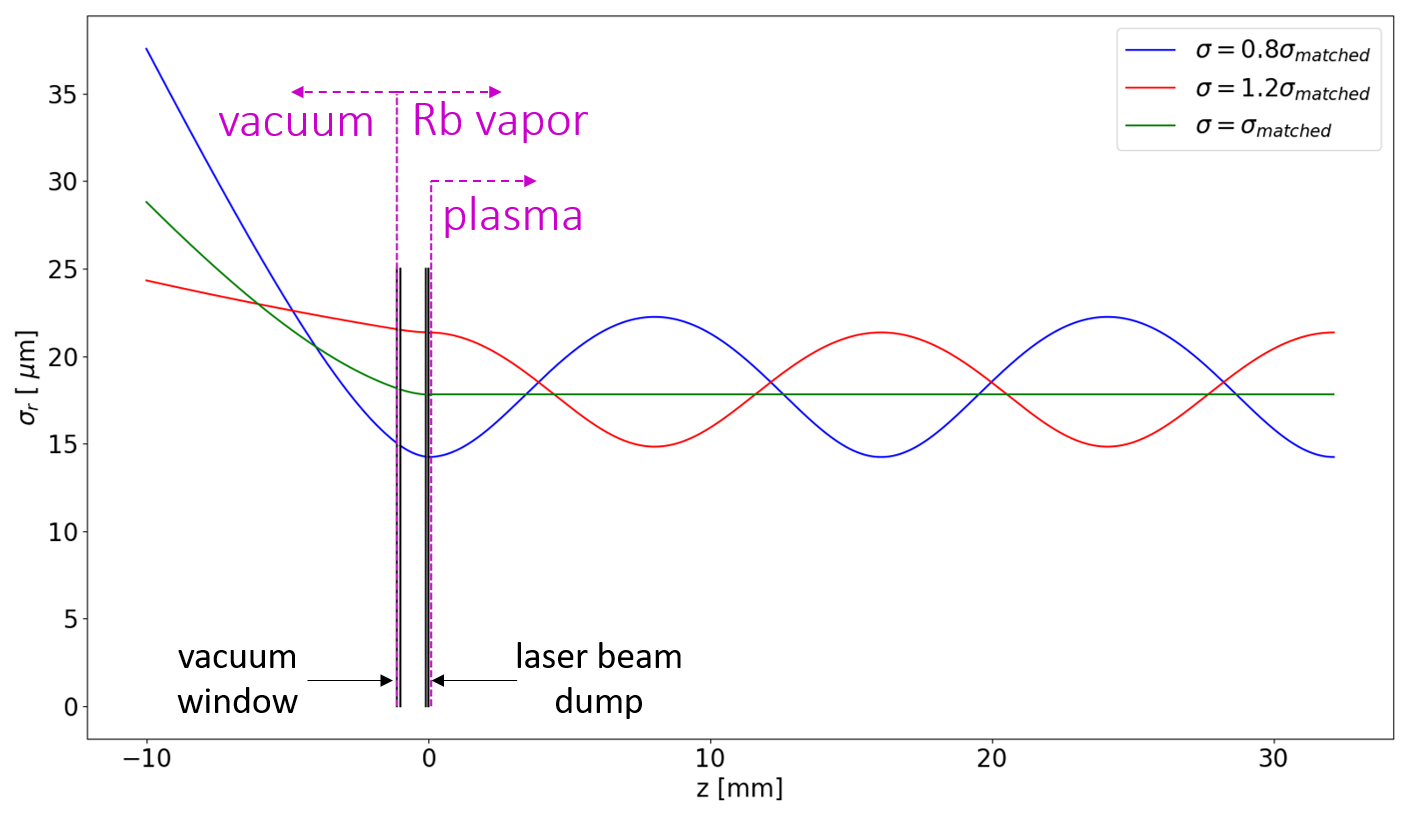}
\caption{\label{fig:betatron}Transverse electron bunch size in vacuum ($z < -1\,\textnormal{mm}$), in Rb vapor between foils ($-1< z < 0 \,\textnormal{mm}$) and along the plasma ($z > 0.1 \,\textnormal{mm}$) for the case of a matched (green line), under-matched (red) and over-matched (blue) beam.}
\end{minipage}\hspace{2pc}%
\begin{minipage}{18pc}
\includegraphics[width=17pc]{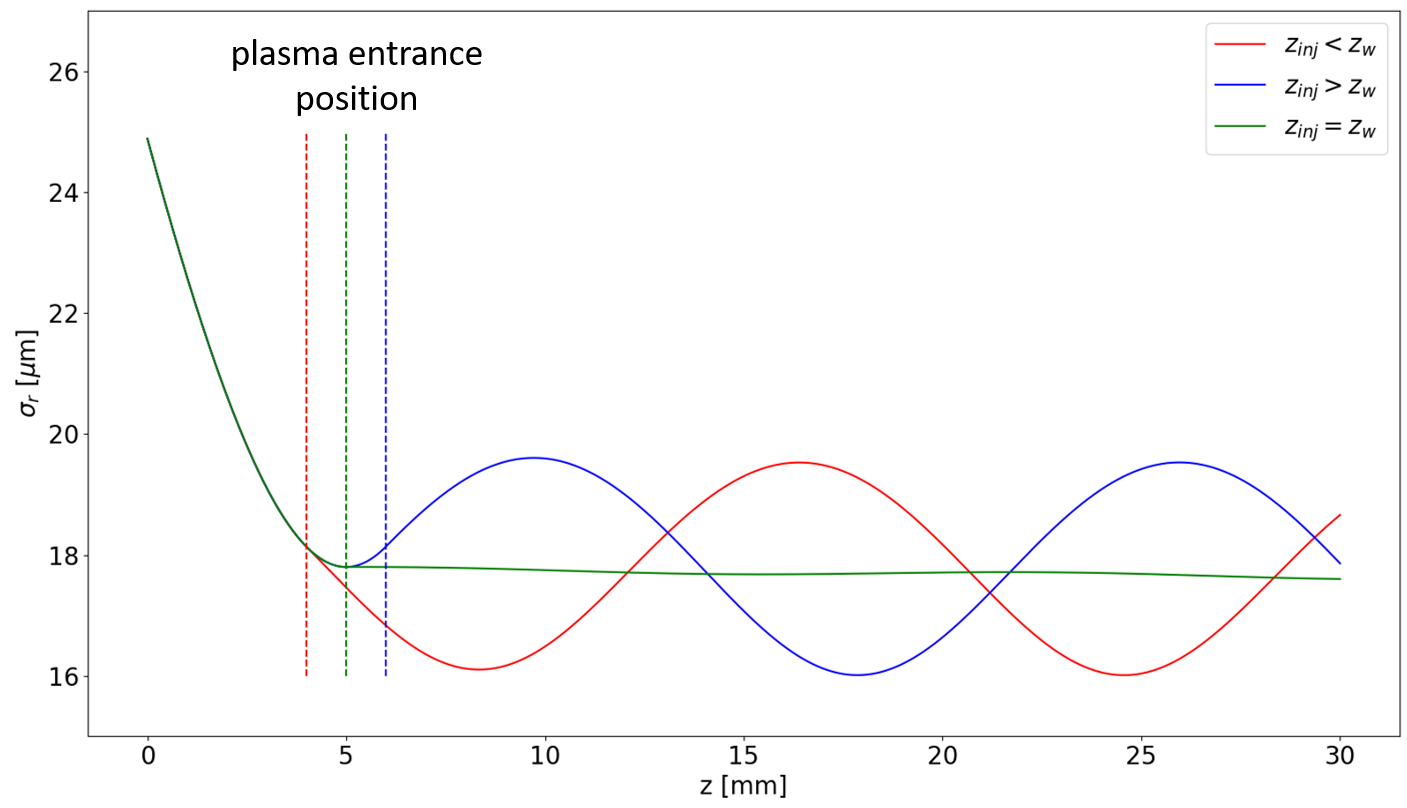}
\caption{\label{fig:betatron_waist}Transverse electron bunch size in the plasma for the case of a beam injected at the waist ({$z_{inj} = z_w$}, green line), of an early injected beam (${z_{inj} = z_w - 1\,\textnormal{mm}}$, red line) and of a late injected beam ({$z_{inj} = z_w + 1\,\textnormal{mm}$}, blue line). $\beta = \beta^{*}$ for all the examples.}
\end{minipage} 
\end{figure}

 When the beam is correctly matched to the plasma ion column (see the green line in Figure $\ref{fig:betatron}$), its transverse size does not oscillate along the plasma.  The bunch charge density does not change and the blowout and beam loading are maintained along the entire plasma length. On the contrary, when the beam is mismatched (see blue and red lines) betatron oscillations of the beam envelope take place. The beam size along the plasma is obtained by integrating Equation $\ref{eq:envelope}$ with particular initial condition:
\begin{equation}
    \sigma (z) = \sqrt{\frac{\epsilon_g (\sqrt{F^2 +1} + Fcos(2\sqrt{K_{\beta}} z))^{1/2}}{\sqrt{K_{\beta}}}},
\end{equation}
where $F = \frac{(K_{\beta}\sigma^{*4} / \epsilon_g^2) -1}{2\sqrt{K_{\beta}}\sigma^{*2} / \epsilon_g}$.
Moreover, even though the betatron function at the waist is equal to $\beta^{*}$ (calculated with Equation $\ref{eq:matching_beta}$), the envelope starts oscillating when the beam is not injected at the waist ($\sigma'(z_{inj}) \ne 0$), as shown in Figure $\ref{fig:betatron_waist}$. Since the plasma entrance is determined by the location of the counter-propagating ionization laser pulse and electron bunch meeting point, mismatched caused by $\sigma'(z_{inj}) \ne 0$ can be corrected by adjusting the relative timing between the pulse and the bunch. The value of the betatron function also determines the required timing precision needed for the meeting point. This time has to be much shorter than the transition time of the bunch over one $\beta^{*}$. In this case, $\beta^{*} / c = 17 \,\textnormal{ps}$. We also note that, as the crossing distance is on the order of $\sigma_z << \beta^{*}$, the entire bunch can be considered as injected and matched at once. Instead, mismatching caused by the wrong transverse waist size can only be corrected by adjusting the incoming beam optical properties.

\subsection{Effect of acceleration on electron beam matching}
\par
\begin{figure}[h!]
\includegraphics[width=18pc]{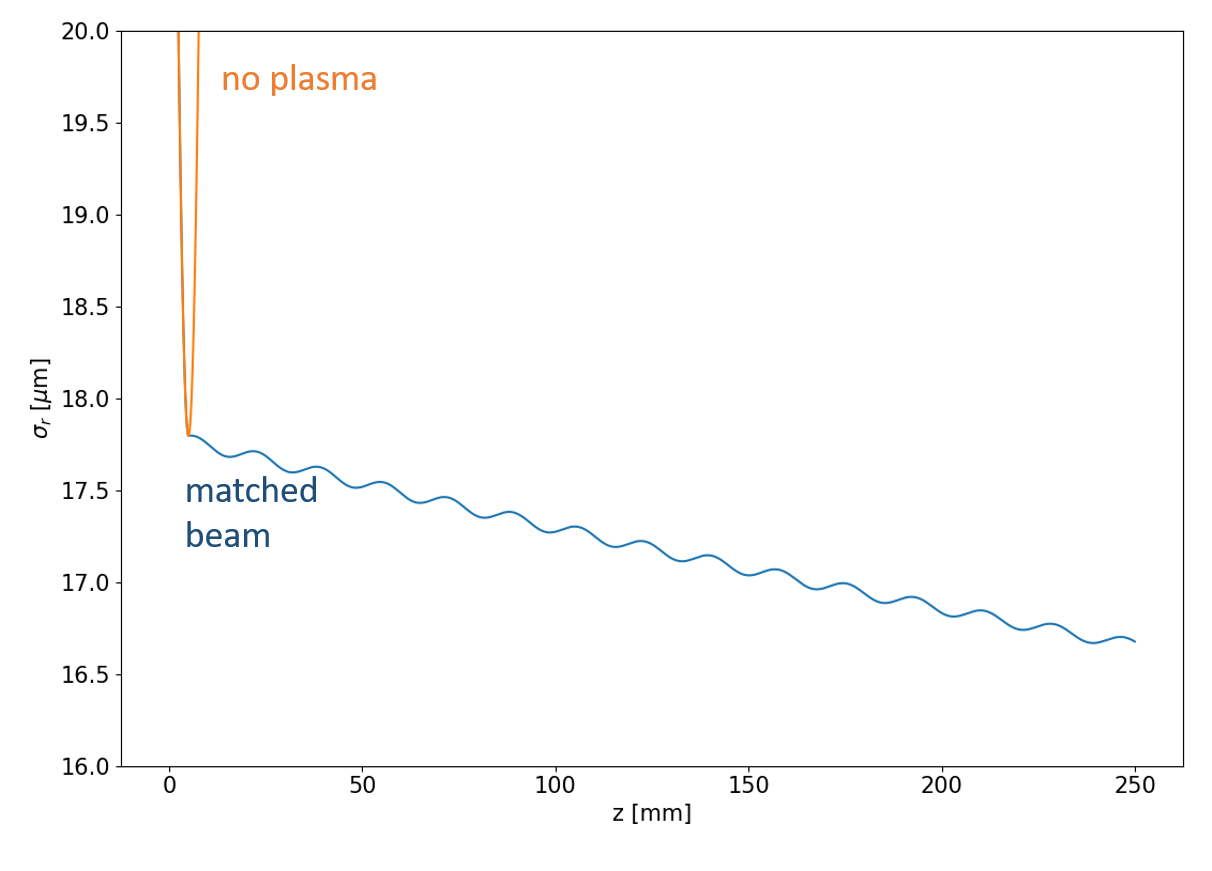}\hspace{2pc}%
\begin{minipage}[b]{16pc}\caption{\label{fig:envelope_acc}Electron beam envelope for a matched beam accelerated in plasma (blue line), and without plasma (orange line)}
\end{minipage}
\end{figure}

 So far, we determined the matching condition at the plasma entrance for the incoming beam energy. But energy gain occurs along the plasma ($\gamma(z)=\gamma_0( 1+ \frac{d\gamma}{dz} dz)$), potentially leading to loss of the matching condition.  However, when the energy gain per unit length is small enough so that $\frac{d\gamma}{dz}<< \frac{\gamma}{\lambda_{\beta}}$, the matching is mostly maintained. The transverse bunch size therefore adiabatically  adjusts to satisfy the matching condition as the energy changes, according to $\sigma \propto \gamma^{1/4}$ (from Equation $\ref{eq:matching_beta}$). In this case, $\gamma = \gamma(z)$ in Equation $\ref{eq:envelope}$, that is solved numerically. We consider a constant energy gain $\frac{d\gamma}{dz} = 200 \,\textnormal{MeV/m}$ (from Run1 experimental results $\cite{NATURE}$), that satisfies the adiabaticity condition, and we assume that the normalized emittance is preserved during the acceleration. Figure $\ref{fig:envelope_acc}$ shows that the bunch size decreases overall, but the approximate matching leads to small envelope oscillations. With the adiabatic matching, the transverse beam size and the betatron function are, after the acceleration over $10\,\textnormal{m}$ of plasma, $\sim 10 \,\textnormal{$\mu$m}$ and $\sim 2 \,\textnormal{cm}$, respectively. This is important for the magnetic energy spectrometer design. We also note here that the effect of the density ramp at the plasma exit must be included. Still, the matching condition (Equation $\ref{eq:matching_beta}$) is defined for only one energy value. Therefore, in order to have the whole beam nearly matched, a small energy spread of the incoming beam, and the beam loading, are necessary conditions. And, to satisfy the request of a constant normalized emittance, the beam must be accelerated in the blowout regime. It is then clear that matching, blowout and beam loading are reciprocal conditions. Blowout and beam loading are achieved through proper charge and length adjustment of the electron bunch (for a given radius, determined from matching to the pure ion column), that is not discussed here. 

%

\section{Conclusions}

AWAKE Run2 focuses on producing a high-energy and high-quality beam suitable for high-energy physics applications. To do so, the Run1 experimental setup is modified in order to precisely control the electron beam injection in the proton-driven plasma wakefields. The emittance preservation and the low final energy spread are achieved exploiting full blowout of the plasma electrons, beam loading of the wakefields, and matching of the electron beam to the plasma ion column focusing force. For a given plasma electron density, we therefore define the necessary incoming electron beam parameters, considering the amount of material that the bunch has to cross before entering the plasma, and we find that the use of foils relax the condition on the incoming electron beam betatron function. The electron bunch parameters for matching are determined for the case of a low electron bunch normalized emittance ($20 \,\textnormal{mm}\cdot\textnormal{mrad}$). Full parameters were determined for the case of a single driver proton microbunch $\cite{VERONICA:BLOWOUT}$ and a witness bunch normalized emittance of $2\,\textnormal{mm}\cdot\textnormal{mrad} $. Parameters for the higher emittance case need to be determined for the fields driven by the self-modulated proton bunch and also for lower densities (e.g. $\textnormal{from } 2 \textnormal{ to }4\cdot 10^{14} \,\textnormal{cm$\textsuperscript{-3}$}$). To unequivocally prove the matching of the electron beam with the plasma wakefields, we will experimentally study the accelerated beam properties as a function of the incoming beam parameters. Therefore, it will be necessary to measure the transverse beam size and position at the entrance of the accelerating plasma section with $\mu$m-resolution. This will require a challenging design and integration of the diagnostics, due to the compact geometry of the system. Once the acceleration of a high-quality electron bunch is successfully proven, the final energy could be increased by simply scaling up the length of the accelerating plasma section.

\section*{References}

\bibliography{mybibfile}

\end{document}